\documentclass[aps,twocolumn,showkeys,groupedaddress]{revtex4-1}
\usepackage{graphicx}
\usepackage{amsmath}
\usepackage{amssymb}
\usepackage{longtable}
\usepackage{dcolumn}
\usepackage{natbib}
\begin{document}

\title{Regularizing binding energy distributions and thermodynamics of hydration. Theory and application to water modeled with classical and {\em ab initio\/} simulations}
\author{Val\'ery Weber}\thanks{Email: vwe@zurich.ibm.com}
\affiliation{IBM Research Division, Zurich Research Laboratory, 8803 Ruschlikon, Switzerland}
\author{Safir Merchant}
\author{D. Asthagiri}\thanks{Email: dilipa@jhu.edu}
\affiliation{Department of Chemical and Biomolecular Engineering, Johns Hopkins University, Baltimore, MD 21218}

\date{\today}
\begin{abstract}
The high-energy tail of the distribution of solute-solvent interaction energies is poorly characterized for condensed systems, but this
tail region is of principal interest in determining the excess free energy of the solute. We introduce external fields centered on the solute to modulate the short-range repulsive interaction between the solute and solvent. This regularizes the binding energy distribution and makes it easy to calculate the free energy of the solute with the field. Together with the work done to apply the field in the presence and absence of the solute, we calculate the excess chemical potential of the solute. We present the formal development of this idea and apply it to study liquid water. 
\end{abstract} 


\maketitle

The excess chemical potential, $\mu^{\rm ex}_{\rm x}$, of a solute ($\rm x$) within a general thermodynamic system (the solvent) is that part of the Gibbs free energy that would vanish if the interaction between the solute and solvent were to vanish. For this reason the excess chemical potential is of most interest in understanding a complex system on a molecular basis. Since the excess chemical potential is measurable, calculating it also serves as an important validation of molecular simulations. 

In this communication, we present a new approach to calculate $\mu^{\rm ex}_{\rm x}$, one that sidesteps the current dominant
paradigm based on alchemically changing solutes. Our approach, a generalization of the quasichemical organization  \cite{lrp:apc02,*lrp:book,*lrp:cpms} of the potential distribution theorem \cite{widom:jpc82}, rests on appreciating and exploiting the different energies with which a solute interacts with a solvent at different spatial scales. The approach leads to a transparent accounting of the hydration thermodynamics and is readily applicable to systems modeled by {\em ab initio\/} potentials or molecularly complex solutes such as proteins. 
In this communication we present our results for liquid water modeled by both empirical and {\em ab initio\/} potentials. Studies on a protein modeled by empirical potentials and of aqueous ions modeled by {\em ab initio\/} potentials will be presented in subsequent articles. 

To appreciate the need for alchemical approaches, consider the formal relation between $\mu^{\rm ex}_{\rm x}$ and solute-solvent interactions \cite{lrp:apc02,lrp:book,lrp:cpms,widom:jpc82}: 
\begin{subequations}
\label{eq:mudefn}
\begin{eqnarray}
\beta \mu^{\rm ex}_{\rm x} & = & \ln \int P_{\rm x}(\varepsilon) e^{\beta \varepsilon} d\varepsilon = \ln \langle e^{\beta\varepsilon} \rangle \label{eq:inv} \\
& = & -\ln \int P^{(0)}_{\rm x}(\varepsilon) e^{-\beta \varepsilon} d\varepsilon  = -\ln \langle e^{-\beta\varepsilon}\rangle_0 \label{eq:for} \, ,
\end{eqnarray}
\end{subequations}
where $\beta = 1/k_{\rm B}T$ and $\varepsilon = U_{N+1} - U_{N} - U_{\rm x}$ is the interaction energy of the solute (x) with the
solvent. $U_{N+1}$ is the potential energy of the $N+1$ particle system comprising the $N$ solvent molecules and the one solute molecule. $U_{N}$ is the potential energy of the solvent and $U_{\rm x}$ is the potential energy of the solute. $P_{\rm x}(\varepsilon)$ is the density distribution of $\varepsilon$ in a system in which the solute and the solvent are thermally coupled, the  averaging under these conditions is denoted by $\langle\ldots\rangle$.  $P_{\rm x}^{(0)}(\varepsilon)$ is the density distribution when the solute and the solvent are thermally uncoupled, with $\langle\ldots\rangle_0$ denoting averaging in this instance.  

For condensed systems the high-$\varepsilon$ tail of $P_{\rm x}(\varepsilon)$ or the low-$\varepsilon$ tail of $P_{\rm x}^{(0)}(\varepsilon)$ are rarely well sampled and thus using either of Eqs.~\ref{eq:mudefn} usually fails. It is for this reason that $\mu^{\rm ex}_{\rm x}$ 
is often calculated by accumulating the work done in slowly transforming a solute from a noninteracting solute to a fully interacting solute. However, such alchemical transformations pose conceptual challenges in the context of {\em ab initio\/} simulations; 
an intermediate solute with fractional charge (and uncertain spin state) can prove troublesome within
quantum chemical calculations.  For example, using {\em ab initio\/} simulations, an alchemical approach was used to calculate the hydration free energy of some cations, but the same approach could not be used for a chloride anion  \cite{rempe:jcp09}. 
Thus the broader applicability of alchemical approaches within {\em ab initio\/} simulations remains unknown, and in particular, we are unaware of a similar study for the hydration free energy of a water molecule. 

In contrast to alchemical transformations, our strategy is to regularize the $P_{\rm x}(\varepsilon)$ distribution by imposing a constraint \cite{safirphd}. Here the constraint is 
an external field $\phi(r)$ centered on the particle such that in the presence of this field the solute-solvent interactions are better behaved.  Using the well-known rule of averages (for example, see \cite{lrp:apc02,lrp:book,lrp:cpms}), 
  \begin{eqnarray}
 \langle F \rangle_{u} = \frac{\langle e^{-\beta u} F \rangle_0}{\langle e^{-\beta u} \rangle_0} \, \nonumber 
 \end{eqnarray}
 for any potential $u$ and the mechanical variable $F$, where $\langle\ldots\rangle_u$ denotes averaging when the solvent evolves in the presence of the additional potential $u$, we can show that 
\begin{eqnarray}
\beta\mu^{\rm ex}_{\rm x} = \ln  \langle e^{-\beta \phi} \rangle -  \ln \langle e^{-\beta \phi} \rangle_0  -  \ln \langle e^{-\beta \varepsilon} \rangle_\phi  \, .
\label{eq:genqc}
\end{eqnarray}
The quantity, $\ln  \langle e^{-\beta \phi} \rangle$, is the negative of the work done to apply $\phi$ in the solute-solvent system; $-  \ln \langle e^{-\beta \phi} \rangle_0$ is the work done to apply $\phi$ in the neat solvent system;  and $-  \ln \langle e^{-\beta \varepsilon} \rangle_\phi$  is the interaction free energy of the uncoupled solute with the solvent in the presence of $\phi$. Notice
that whereas $-  \ln \langle e^{-\beta \varepsilon} \rangle_0$ can prove challenging to estimate because of close solute-solvent contacts, $-  \ln \langle e^{-\beta \varepsilon} \rangle_\phi$ is expected to be more amenable to a direct estimation. Indeed, if $\phi$ is chosen such that the regularized binding energy distribution $P^{(0)}_{\rm x}(\varepsilon | \phi)$ is a Gaussian of mean $\langle \varepsilon \rangle_{\phi}$ and variance $\langle \delta \varepsilon^2 \rangle_\phi$, where $\delta \varepsilon = \varepsilon - \langle\varepsilon\rangle_\phi$, we have
\begin{eqnarray}
-  \ln \langle e^{-\beta \varepsilon} \rangle_\phi = \beta \langle \varepsilon\rangle_\phi - \frac{\beta^2}{2} \langle\delta\varepsilon^2\rangle_\phi \, .
\label{eq:gauss}
\end{eqnarray}
The first two terms on the right-hand side of Eq.~\ref{eq:genqc} can be obtained using  thermodynamic integration, but 
the solute-solvent Hamiltonian is unchanged, an obvious advantage in {\em ab initio\/} simulations.  For example, 
if $\phi(\xi)$ varies between zero to its final value as the parameter $\xi$ is varied from 
zero to $\lambda$, then \cite{darve:cpms}
\begin{eqnarray}
-  \ln \langle e^{-\beta \phi} \rangle_0 = \beta\int_{0}^{\lambda} \left\langle \frac{\partial\phi}{\partial \xi}\right\rangle_{\phi} d\xi \, ,
\label{eq:ti}
\end{eqnarray}
where the ensemble averaging is in the presence of the field at the current value of $\xi$. 

Eq.~\ref{eq:genqc} is a generalization of the quasichemical approach~\cite{lrp:apc02,lrp:book,lrp:cpms} for any external field $\phi$.   In the original quasichemical development \cite{lrp:apc02,lrp:book,lrp:cpms}, solvent molecules were strictly excluded from a sphere of radius $\lambda$ around the solute. In the present notation, this amounts to applying the field   
\begin{eqnarray}
\phi_{\rm h}(r; \lambda) = \begin{cases}
\infty & r < \lambda \\
0 &  r \geq \lambda 
\end{cases}
\label{eq:hard}
\end{eqnarray}
where $r$ is the distance between the solvent and the solute. $x_0 = \langle e^{-\beta \phi_{\rm h}}\rangle$ is the probability of observing zero solvent molecules in a coordination sphere of radius $\lambda$ around the solute,  $p_0 = \langle e^{-\beta \phi_{\rm h}}\rangle_0$ is the probability of observing an empty cavity of radius $\lambda$ in neat water, and $\beta\mu^{\rm ex}_{\rm outer} =  -  \ln \langle e^{-\beta \varepsilon} \rangle_{\phi_{\rm h}}$ is the contribution to the free energy due to the interaction of the solute with the solvent outside the coordination sphere.  For solutes that interact strongly with water or for solutes with complicated molecular shape, obtaining 
$x_0$ and $p_0$ can prove difficult, especially for a  large cavity radius \cite{asthagiri:jacs07}.  The flexibility to choose any field and the thermodynamic integration method (Eq.~\ref{eq:ti}) both alleviate this difficulty within the context of Eq.~\ref{eq:genqc}.  In analogy to the original quasichemical notation, since the fields used here (Eqs.~\ref{eq:phi}) model soft-cavities, we rewrite Eq.~\ref{eq:genqc} as 
\begin{eqnarray}
\beta\mu^{\rm ex}_{\rm w} = \ln  x_s -  \ln p_s  + \beta \mu^{\rm ex}_{\rm outer, s} \, ,
\label{eq:genqc1}
\end{eqnarray}
where the solute (x) is a distinguished water (w) molecule. 

In this study we tested two fields
\begin{subequations}
\label{eq:phi}
\begin{eqnarray}
\phi_{\rm ramp}(r; \lambda) & = & a \left( \sqrt{(r-\lambda)^2 + b^2} - b\right) \label{eq:ramp} \\
\phi_{\rm lj}(r; \lambda) & = & 4a\bigg[ \left(\frac{b}{r-\lambda+\sqrt[6]{2}b}\right)^{12} \nonumber \\
& & - \left(\frac{b}{r-\lambda+\sqrt[6]{2}b}\right)^{6} \bigg] + a \label{eq:lj} \; ,
\end{eqnarray}
\end{subequations}
where $a$ and $b$ are positive constants and $r < \lambda$;  $\phi(r; \lambda) = 0$ for $r \geq \lambda$. 

Before presenting the simulation details, we note that a variant of the quasichemical approach with a soft-cutoff has been presented recently  \cite{lrp:softcutoff}. That approach uses a probabilistic model to partition solvent between the inner shell ($r < \lambda$) and the bulk and is thus rather different from the approach here.  

We next consider the simulation implementation of Eq.~\ref{eq:genqc}. 

{\bf Classical simulations:} Classical simulations were performed with the NAMD code \cite{namd}. Using the Tcl-interface in NAMD, the force $\partial \phi/\partial \lambda$ is applied to the solvent molecules within $\lambda$ and the ensemble average
(integrand in Eq.~\ref{eq:ti}) estimated. From the average force-vs-$\lambda$ data, the free energies are constructed. 

We simulate a cubic box of 512 SPC/E \cite{spce} water molecules at a temperature of 298~K using a Langevin thermostat and a pressure of 1~bar using a Langevin barostat \cite{feller:jcp95}. The decay constant for the thermostat was 1 ps$^{-1}$. The barostat piston period was 200~fs and the decay time was 100~fs. The SHAKE \cite{md:shake} algorithm was used to constrain the geometry of each water molecule. The solute water molecule was fixed at the center of the simulation cell. 

We sample  $\lambda$ between 0 and 5.0~{\AA}. First the $\lambda = 2.5$~{\AA} state was equilibrated for 
2.4~ns and then all other $\lambda$ values were generated by successively changing $\lambda$ by 
$\pm 0.1$~{\AA}.  At each $\lambda$, we performed simulations for a total of 600~ps and used the last 500~ps for analysis. 

For calculating $\mu^{\rm ex}_{\rm outer,s}$, we extend the trajectory for states with $\lambda$ between 4.5~{\AA} and 5.0~{\AA} by an additional 600~ps, saving configurations every 0.5~ps. For the last 1100 frames, we inserted a test water molecule in the center of the cavity and  calculated its binding energy with the medium. Five randomly chosen orientations of the test water were used per frame. A direct 
numerical estimate of $ \langle e^{-\beta \varepsilon} \rangle_\phi$ and modeling the binding energy as a Gaussian gave nearly the same values for $\mu^{\rm ex}_{\rm outer, s}$. 

{\bf Ab initio simulations:} We simulated water using the BLYP-D electron density functional and the {\sc cp2k} code \cite{cp2knew}. 
The parameters for the electronic structure calculation are as in earlier studies \cite{weber:jcp10a,weber:jcp10b}.  In contrast to the specification of the external force $\partial \phi/\partial \lambda$ in NAMD, the external potential $\phi$ is provided as an input to the {\sc cp2k} code. Since the code calculates the force by differentiating an interpolation function of the potential, we use a non-zero $b$ parameter in $\phi_{\rm ramp}$ (Eq.~\ref{eq:ramp}) to ensure forces are well-behaved at the cavity boundary. 

We simulate the liquid at a density of 0.997 g/cm$^3$ (number density of 33.33 nm$^{-3}$) and a temperature of 350 K. The system contains 64 water molecules. To sample the NVT ensemble, we use the hybrid Monte Carlo method~\cite{weber:jcp10a}. The initial configuration is taken from a previous MD simulation with the same functional~\cite{weber:jcp10b}. During  the first 500 sweeps, the time step for integrating the equations of motion was adjusted to provide an acceptance ratio of 70\%. Then the time step was fixed and the system equilibrated for another 500 sweeps. (One sweep of the HMC comprises 50 molecular dynamics time steps.) 

For estimating $\ln x_s$, a chain of HMC simulation starting at $\lambda = 2.5$ {\AA} and going up to $3.75$~{\AA} in increments
of 0.25 {\AA} was performed. During the construction of the chain, at each $\lambda$ we performed 300 sweeps. 
Subsequently, for calculating averages, simulations at each $\lambda$ were extended for another 1200 sweeps. 
For estimating $\ln p_s$, a similar strategy is used but with $\lambda = 1.75$ {\AA} to $\lambda = 3.75$ {\AA}.  
For both $\ln x_s$ and $\ln p_s$, from the total of 1500 sweeps per $\lambda$, we use the last 1000 for analysis.

To compute $\mu^{\rm ex}_{\rm outer,s}$, we chose the last 1000 configurations. We estimate the binding energy distribution using the procedure documented earlier  \cite{weber:jcp10b}, except that we perform 16 test particle insertions per configuration. $\mu^{\rm ex}_{\rm outer,s}$ was estimated using the ensemble average $ \langle e^{-\beta \varepsilon} \rangle_\phi$ and by the Gaussian model. 

We next present the results and their discussion. 

\begin{figure}[h!]
\includegraphics{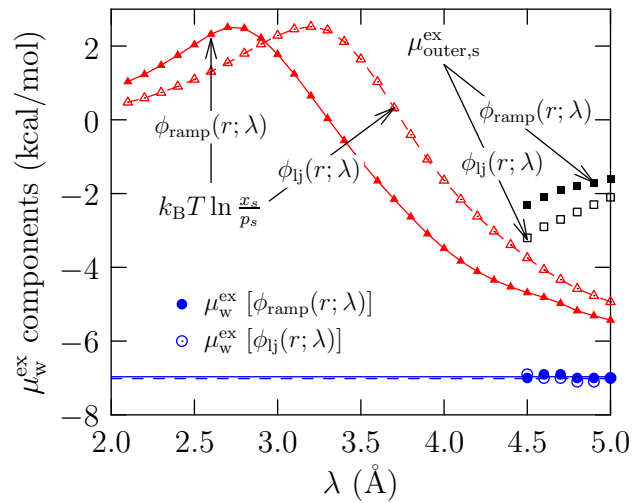}
\caption{Various components of the free energy obtained using Eq.~\ref{eq:genqc1} ($\rm w \equiv H_2O$) for the SPC/E water model. $k_{\rm B}T\ln x_s / p_s$ smoothly approaches zero (0) as $\lambda \rightarrow 0$. \underline{Open symbols}:  $\phi_{\rm lj}$ (Eq.~\ref{eq:lj}) with $a = 0.155$~kcal/mol and $b=3.1655$~{\AA}, the LJ energy and diameter parameters for SPC/E water.  
\underline{Filled symbols}:  $\phi_{\rm ramp}$ (Eq.~\ref{eq:ramp}) with $b=0$ and $a = 5.0$~kcal/mol. 
The average value of $\mu^{\rm ex}_{\rm w}$ over $\lambda$ is shown by the solid blue ($\phi_{\rm ramp}$) and dashed blue ($\phi_{\rm lj}$) lines;  $\mu^{\rm ex}_{\rm w} \approx -7.0$~kcal/mol is in excellent agreement with estimates based on histogram overlap ($-6.85$~kcal/mol, Ref.\ \onlinecite{merchant:jcp11a}) and from the equilibrium vapor and liquid densities ($-7.0$~kcal/mol, Ref.\ \onlinecite{lrp:jpcb09}). At $\lambda = 5.0$~{\AA}, the statistical uncertainty in  $k_{\rm B}T \ln (x_s/p_s)$ is  0.1~kcal/mol. The uncertainty in $\mu^{\rm ex}_{\rm outer, s}$ can be ignored.}\label{fg:spcewater}
\end{figure}

{\bf Classical Simulations}: Figure~\ref{fg:spcewater} collects the results for the SPC/E water simulation for different choices of the external field. 

As expected, the net chemical potential is nearly independent of $\lambda$ or the field, as it obviously must be. 

From the rule of averages, we can show that 
\begin{eqnarray}
p_0(r_{\rm h}) =  p_s (\lambda) \cdot f(r_{\rm h})\cdot \langle e^{\beta\phi}| r \geq r_{\rm h} \rangle_\phi
\label{eq:muhs}
\end{eqnarray}
where $p_0(r_{\rm h})$ is the probability to observe a hard-sphere of radius $r_{\rm h}$, $f(r_{\rm h})$ 
is the fraction of configurations where a hard-sphere of radius $r_{\rm h}$ is found in a simulation with an external field $\phi(r; \lambda)$ and $\langle \ldots | r \geq r_{\rm h}\rangle_\phi$ denotes ensemble averaging over such configurations. The location of water molecules is defined by $r$ (Eq.~\ref{eq:phi}) and $\lambda$ specifies the range up to which the field is active. (Here  $\lambda \geq r_{\rm h}$.) 
\begin{figure}[h!]
\includegraphics{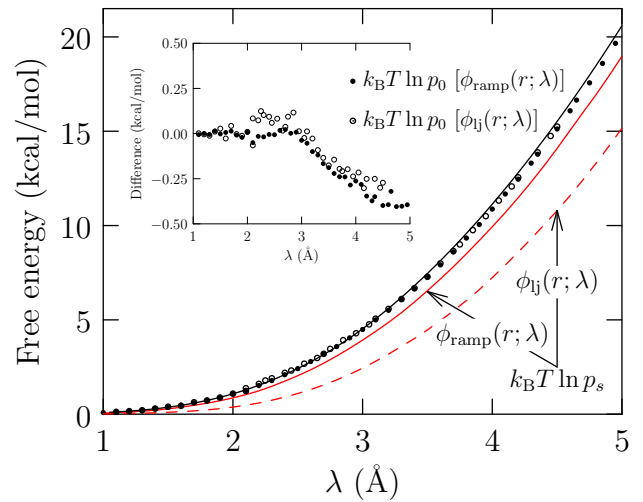}
\caption{The free energy to create empty soft-and hard-cavities of radius $\lambda$ in neat water. \underline{Red lines}: soft cavities. 
\underline{Filled symbols}: hard-cavity from $\phi_{\rm ramp}$ (Eq.~\ref{eq:ramp}).  \underline{Open Symbols}: hard-cavity from $\phi_{\rm lj}$ (Eq.~\ref{eq:lj}). Parameters for the $\phi$'s are noted in Fig.~\ref{fg:spcewater}.  The inset shows the deviation of the calculated $-k_{\rm B}T \ln p_0$ from the revised scaled particle theory~\cite{Ashbaugh:rmp} prediction (solid black line, main figure).}\label{fg:muhs} 
\end{figure}
Figure~\ref{fg:muhs} shows the results of such a calculation,  with $r_{\rm h}$ chosen such that $f(r_{\rm h})$ 
is around 0.25 for a given $\lambda$. As the figure shows, using either field (Eqs.~\ref{eq:phi}) we predict the same value of $p_0 (r_{\rm h})$ within statistical uncertainties of about 0.2~kcal/mol.
 
Fig.~\ref{fg:muhs} reveals that for $\lambda > 3.0$~{\AA}, the calculated $p_0$ is systematically slightly larger (the deviation in $-k_{\rm B}T \ln p_0$ is negative) relative to the revised scaled particle theory (SPT) \cite{Ashbaugh:rmp} predictions.  This small difference 
is likely due to a lower bulk surface tension predicted by the SPC/E model \cite{vega:jcpst}, whereas the 
SPT predictions are based on the experimental bulk surface tension. Indeed, using the surface tension of SPC/E, we are able to 
reparametrize SPT to reproduce the $p_0(r_h)$ obtained in this study. 

{\bf Ab initio Simulations}: Table~\ref{tb:mu} collects the estimated $\mu^{\rm ex}_{\rm w}$ of water. Using $\phi_{\rm ramp}$
(Eq.~\ref{eq:ramp}), the average of $\mu^{\rm ex}_{\rm w}$ over the $\lambda$ values considered here is about $-6.9\pm0.4$~kcal/mol). Earlier, using a hard-sphere conditioning and also an order of magnitude more data --- 110k energy values~\cite{weber:jcp10b} versus 16k energy values used here --- for calculating $\mu^{\rm ex}_{\rm outer, s}$ we found $\mu^{\rm ex}_{\rm w} \approx -6.0$~kcal/mol for a 64-water molecule system at an average temperature of 362~K \cite{weber:jcp10b}.  The present estimate appears in fair agreement with the earlier result, especially considering the difference in temperatures and also the amount of data collected. 
\begin{table}
 \centering
  \caption{The different contributions to the excess chemical potential of water obtained from the {\it ab initio} HMC simulations.  
  $\mu_{outer,s}^{ex}$ is calculated as an ensemble average; the result from a Gaussian fit (Fig.~\ref{fg:pe}) to the binding energy data is given in parenthesis.   Eq.~\ref{eq:ramp} with $b = 0.001$~{\AA} and  $a = 10$~kcal/mol is used to regularize the interactions. The first three rows are using Eq.~\ref{eq:ramp} and the next three rows are with Eq.~\ref{eq:lj}.  
   Energies are in kcal/mol. Uncertainties are at 1$\sigma$ level. }\label{tb:mu}
  \begin{ruledtabular}
  \begin{tabular}{ccccc}
    $\lambda$ &$k_{\rm B} T\ln x_s$ & $-k_{\rm B}T\ln p_s$ & $\mu_{\rm outer,s}^{ex}$ & $\mu^{ex}_{\rm w}$ \\
    \hline 
    3.25    & $-7.1\pm 0.4$ &  $6.8\pm 0.4$ & $-6.8\pm0.4$ ($-$5.8) & $-7.1 \pm 0.7$ \\
    3.50    &  $-11.5\pm 0.4$ &  $9.5\pm 0.5$ & $-4.3\pm0.2$ ($-$3.8) & $-6.3\pm 0.7$\\
    3.75    &   $-16.3\pm 0.5$ & $11.7\pm 0.5$ & $-2.6\pm0.1$ ($-$2.6) & $-7.2\pm 0.7$ \\ \hline
    3.75    &   $-5.3\pm 0.2$ & $6.5\pm 0.3$ & $-5.0 \pm0.2$ ($-5.3$)  & $-3.8\pm0.4$ \\ 
    4.00    &   $-9.1\pm 0.3$ & $8.6\pm 0.3$ & $-4.7 \pm0.2$ ($-4.4$)  &  $ -5.2\pm 0.5$\\ 
    4.25    &   $-13.4\pm 0.4$ & $11.1\pm 0.3$ & $-3.4 \pm0.1$  ($-3.2$) & $-5.7\pm 0.5$ \\ 
 \end{tabular}
 \end{ruledtabular}
\end{table}

The average $\mu^{\rm ex}_{\rm w}$ using $\phi_{\rm LJ}$ (Eq.~\ref{eq:lj}) is $-5.0\pm 0.4$. The estimates using $\phi_{\rm ramp}$ (Eq.~\ref{eq:ramp}) and $\phi_{\rm LJ}$ (Eq.~\ref{eq:lj}) thus bound the earlier estimate of $-6.0$~kcal/mol \cite{weber:jcp10b}, but in contrast 
to results with the classical potential model (Fig.~\ref{fg:spcewater}), the variation in the numerical estimate of $\mu^{\rm ex}_{\rm w}$ with both $\lambda$ and the field is high (being about 1-2~kcal/mol).  Comparing $k_{\rm B}T \ln x_s$ and $k_{\rm B}T \ln p_s$ 
between $\phi_{\rm ramp}$ and $\phi_{\rm lj}$ (Table~\ref{tb:mu}) suggests that the (large) step-size of 0.25~{\AA}
for integration underlies the observed variation in $\mu^{\rm ex}_{\rm w}$. Identifying optimal integration strategies, perhaps
including non-uniform sampling of $\lambda$, is left for future work. 

\begin{figure}
\includegraphics{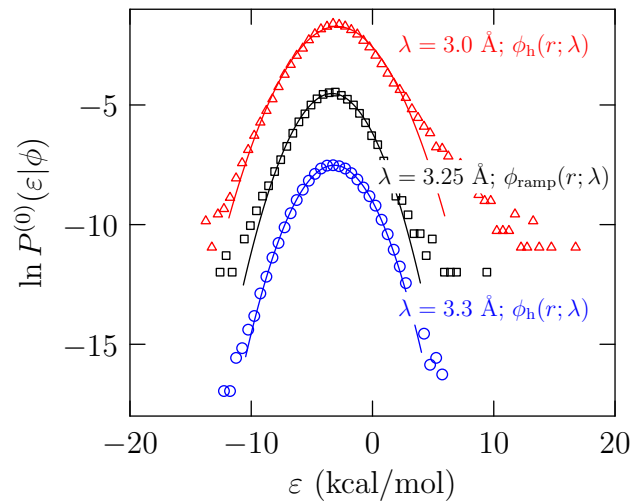}
\caption{Distribution of the interaction energies $P^{(0)}(\varepsilon|\phi)$ of the test particle in the presence of 
different regularizing fields~$\phi$. The binding energies are translated vertically for clarity. \underline{Triangles (red) and circles (blue)}: $\phi_{\rm h}$ is used; data has been taken from our earlier study \cite{weber:jcp10b}. 
\underline{Squares (black)}: $\phi_{\rm ramp}$ with $a = 10$~kcal/mol and $b= 0.001$~{\AA}.   The solid lines are Gaussian fit to the respective distribution. For the radii range considered here, regularizing with Eq.~\ref{eq:ramp} leads to a less well-characterized low-$\varepsilon$ tail relative to Eq.~\ref{eq:hard}, as is expected.}\label{fg:pe}
\end{figure}

Figure~\ref{fg:pe} compares the binding energy of the test particle with the bulk medium for
different choices of the field.
  The value of $\mu^{\rm ex}_{\rm outer, s}$ calculated using a hard-sphere conditioning of
radius 3.0~{\AA}  (see \cite{weber:jcp10b}) is similar  to that for conditioning with $\phi_{\rm ramp}$ ($\lambda = 3.25$~{\AA}); 
not surprisingly, the low-$\varepsilon$ tail is also similar for $\phi_{\rm h}$ ($\lambda=3.0$~{\AA}) and $\phi_{\rm ramp}$ ($\lambda =3.3$~{\AA}). 

In summary, we have presented a method of obtaining the hydration thermodynamics of the solute that avoids alchemical transformation of the solute.  This method allows one to readily calculate the hydration thermodynamics of solutes using {\em ab initio\/} simulations.
Within the quasichemical approach, using a hard-sphere cutoff to demarcate local and long-range interactions restricts one
to small (typically $\leq 4.0$~{\AA}) cavity sizes, since these are the cavities that can be observed in simulations 
with sufficient statistical precision. The present development removes this limitation, and it is now possible  to probe cavity 
sizes that can accommodate small globular proteins (Weber and Asthagiri, in preparation). Thus exploring
the hydration thermodynamics of proteins in aqueous media with the conceptual clarity afforded by the quasichemical approach 
appears to be within reach. Finally, the perspective the regularization approach provides on theories of liquids also remains to be explored. 

\section*{acknowledgments}
The authors warmly thank Claude Daul (University of Fribourg) for computer resources. We thank Hank Ashbaugh (Tulane) for helpful insights about the deviation between our calculations and the scaled particle results. D.A. thanks the donors of the American Chemical Society Petroleum Research Fund for financial support. This research used resources of the National Energy Research Scientific Computing Center, which is supported by the Office of Science of the U.S. Department of Energy under Contract No. DE- AC02-05CH11231.

\newpage


%

\end{document}